\documentclass[prl,twocolumn]{revtex4}
\usepackage{epsfig}
\usepackage{bm}
\usepackage{amsmath}

\begin{document}

\title{Cross-helicity and extended inertial range in MHD turbulence}

\author{A. Bershadskii}

\affiliation{
ICAR, P.O. Box 31155, Jerusalem 91000, Israel
}

\begin{abstract}

  An extended inertial range dominated by the cross-helicity effects has been studied for forced (statistically steady) and for freely decaying magnetohydrodynamic MHD turbulence (with and without imposed/mean magnetic field) using the spatio-temporal distributed chaos approach. Good agreement with results of direct numerical simulations, laboratory measurements in MHD wind (plasma) tunnel, measurements in the Earth's magnetosheath and in the solar wind has been established. A spontaneous breaking of local reflection (mirror) symmetry has been briefly discussed for the MHD turbulence with zero average cross-helicity.  

\end{abstract}

\maketitle

\section{I. Introduction}

  It is now recognized that cross-helicity
$$
H_{cr} = \int {\bf v}({\bf x},t)\cdot {\bf B}({\bf x},t) d{\bf r}  \eqno{(1)}
$$
plays important role in real MHD turbulence: in laboratory experiments, in atmosphere, in solar wind and in solar physics \cite{dmv}-\cite{pou1} (the cross-helicity is usually considered as a measure of relative contribution of the Alfv\'enic waves). Even for the cases with zero (or negligible) average cross-helicity 
$$
h_{cr} = \langle  {\bf v}({\bf x},t)\cdot {\bf B}({\bf x},t) \rangle   \eqno{(2)}    
$$
spatially localised cross-helicity density $({\bf v}({\bf x},t)\cdot {\bf B}({\bf x},t))$ can be rather large \cite{mene},\cite{matt} and, as it  will be shown below, can also play important role in the global MHD dynamics. Since the non-zero cross-helicity is naturally related to lack of the reflection symmetry (unlike the velocity field ${\bf v}$, which is a polar vector, the magnetic field ${\bf B}$ is an axial vector) the last phenomenon represents a kind of the spontaneous breaking of the reflection symmetry.\\

 The role of the cross-helicity is related to the fact that the cross-helicity is an invariant of the non-dissipative (ideal) MHD dynamics. It will be shown below that chaotic (coherent) nature of the MHD dynamics results in an extended inertial range dominated by the average cross-helicity Eq. (2) or by a second-order moment of cross-helicity fluctuations, which is also an invariant of the ideal MHD dynamics (MHD analogue of the Levich-Tsinober invariant \cite{lt},\cite{fl},\cite{l}) and can have a finite non-zero value even when the average cross-helicity is equal to zero. The extended inertial range has been described in the terms of distributed chaos approach and penetrates into near dissipation range of scales.

\section{II. Cross-helicity vs. magnetic energy}

   At the onset of turbulence in plasmas and  fluid dynamics deterministic chaos is often related the exponential power spectra \cite{mm},\cite{kds}
$$
E(k) \propto \exp-(k/k_c)  \eqno{(3)}
$$       
where $k$ is wavenumber and $k_c=$ constant. Figure 1, for instance, shows (in the log-log scales) magnetic energy spectrum obtained in a recent direct numerical simulation (DNS) of the isotropic homogeneous MHD turbulence with a large-scale {\it deterministic} injection (forcing) of kinetic energy (see next section for more detail description). The spectral data used for the Fig. 1 were taken from Fig. 3 of the Ref. \cite{step}. The dashed curve indicates correspondence to the exponential spectrum Eq. (3) and the dotted arrow indicates position of the scale $k_c$. \\ 

\begin{figure} \vspace{-0.7cm}\centering
\epsfig{width=.45\textwidth,file=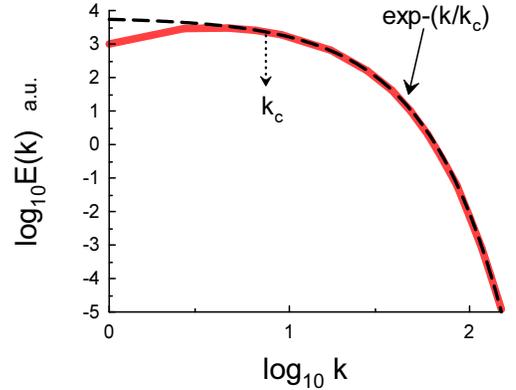} \vspace{-3.8cm}
\caption{ Magnetic energy spectrum for the DNS with {\it deterministic} kinetic energy injection (forcing). } 
\end{figure}

      For a random forcing, however, the parameter $k_c$ in the Eq. (3) fluctuates and in order to compute the spectra we need in ensemble averaging
$$
E(k) \propto \int_0^{\infty} P(k_c) \exp -(k/k_c)dk_c \propto \exp-(k/k_{\beta})^{\beta} \eqno{(4)}
$$    
with certain probability distribution $P(k_c)$. The stretched exponential spectrum in the Eq. (4) is a natural generalization of the exponential spectrum Eq. (3). \\

  One can estimate the probability distribution $P(k_c)$ for large $k_c$ from the Eq. (4) \cite{jon}
$$
P(k_c) \propto k_c^{-1 + \beta/[2(1-\beta)]}~\exp(-\gamma k_c^{\beta/(1-\beta)}) \eqno{(5)}
$$     
  
  Let us consider a scaling relationship between the parameter $k_c$ and characteristic magnetic field strength $B_c$
$$
B_c \propto  k_c^{\alpha}   \eqno{(6)}
$$

 In the case when $B_c$ has Gaussian distribution (with zero mean) a relationship between $\beta$ and $\alpha$
$$
\beta = \frac{2\alpha}{1+2\alpha}  \eqno{(7)}
$$
can be readily obtained from the Eqs. (5-6). 

   In order to obtain value of $\alpha$ (and, consequently, value of $\beta$) let us use the dimensional considerations for the inertial range of scales. Namely
$$
B_c \propto    |\langle  {\bf v}\cdot {\bf B} \rangle|~\varepsilon^{-1/3}~ k_c^{1/3} \eqno{(8)} 
$$
where $\varepsilon$ is the total energy (kinetic plus magnetic) dissipation rate (cf. the Corrsin-Obukhov approach for scaling of passive scalar spectrum Ref. \cite{my} and also Ref. \cite{bs}) . In the inertial range of scales the average cross-helicity $h_{cr} =  \langle  {\bf v}\cdot {\bf B} \rangle$ and the $\varepsilon$ can be considered as adiabatic invariants. Since the $\alpha = 1/3$ in this case we obtain from the Eq. (7) $\beta = 2/5$ i.e.
$$
E(k) \propto \exp-(k/k_{\beta})^{2/5}   \eqno{(9)}
$$  
  
\section{III. Direct numerical simulations - I}

 Dynamics of an electrically conducting incompressible fluid can be described by the MHD equations
$$
 \frac{\partial {\bf v}}{\partial t} = - {\bf v} \cdot \nabla {\bf v} 
    -\frac{1}{\rho} \nabla {\cal P} - [{\bf b} \times (\nabla \times {\bf b})] + \nu \nabla^2  {\bf v} + {\bf f_v} \eqno{(10)}
$$
$$
\frac{\partial {\bf b}}{\partial t} = \nabla \times ( {\bf v} \times
    {\bf b}) +\eta \nabla^2 {\bf b} +  {\bf f_b} \eqno{(11)} 
$$
$$ 
\nabla \cdot {\bf v}=0, ~~~~~~~~~~~\nabla \cdot {\bf b}=0,  \eqno{(12)}
$$
where the velocity field ${\bf v}$ and normalized magnetic field ${\bf b} = {\bf B}/\sqrt{\mu_0\rho}$ have the same dimension (the so-called Alfv\'enic units), $ {\bf f}_v$ and ${\bf f}_b$ are forcing functions of the velocity and magnetic field respectively. 
\begin{figure} \vspace{-2cm}\centering
\epsfig{width=.45\textwidth,file=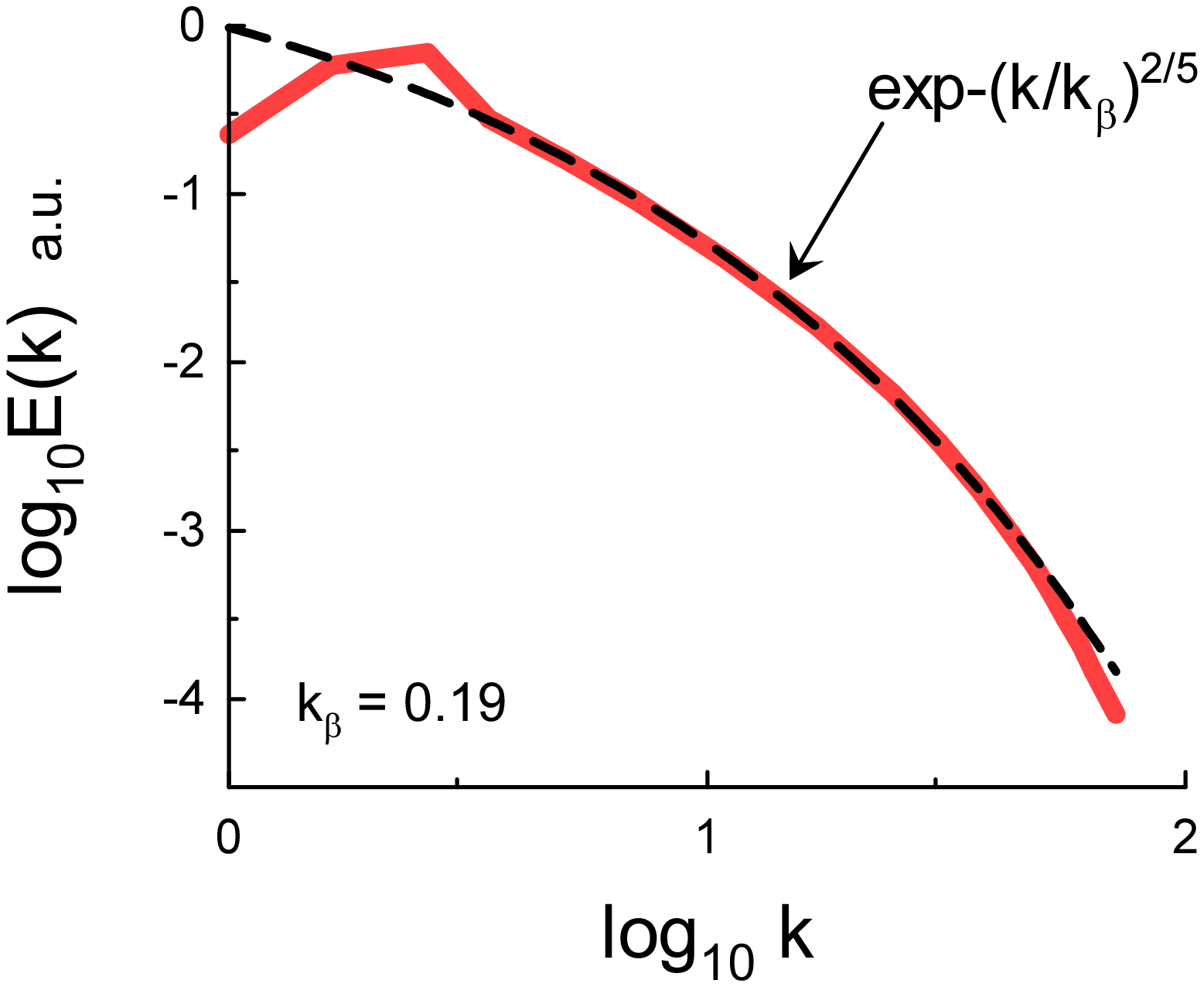} \vspace{-4.2cm}
\caption{ Magnetic energy spectrum for $C=0.3$. } 
\end{figure}
\begin{figure} \vspace{-0.4cm}\centering
\epsfig{width=.45\textwidth,file=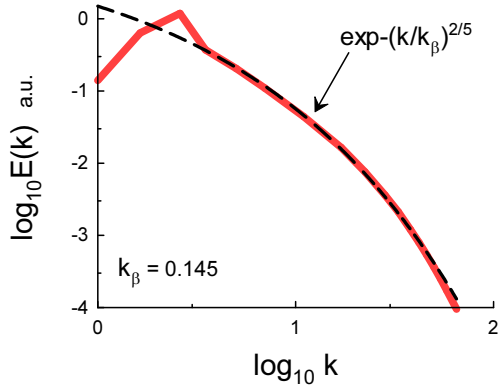} \vspace{-4.3cm}
\caption{Magnetic energy spectrum for $C=0.6$.} 
\end{figure}

  In the above mentioned DNS (Fig. 1) a statistically steady homogeneous and isotropic MHD-turbulence was simulated in a cubic volume using periodic boundary conditions. The {\it deterministic} large-scale forcing terms ${\bf f_b} =0$ and 
$$
{\bf f_v} = \frac{ {\bf v}~}{|{\bf v}|^2},  \eqno{(13)}
$$
were applied in the wavenumber range $2 \leq k \leq 4$. The dissipative parameters were $\nu =0.008$ and $\eta = 0.004$.\\ 

   In the DNS reported in recent papers Refs. \cite{tit},\cite{mag} the deterministic forcing was replaced by a random one (with a controlled level of cross-helicity), which in the Fourier space can be written as
$$
{\bf f_v} ({\bf k}) = (\varepsilon_{s} (1-C)/2)^{1/2} {\bf e_v} ({\bf k}) + (\varepsilon_{s}C/2)^{1/2} {\bf e}_{cr} ({\bf k}),   \eqno{(14)}
$$
$$
{\bf f_b} ({\bf k}) = (\varepsilon_{s} (1-C)/2)^{1/2} {\bf e_b} ({\bf k}) + (\varepsilon_{s}C/2)^{1/2} {\bf e}_{cr} ({\bf k}),   \eqno{(15)}
$$
where 
$$
{\bf e_{v,b,cr}}({\bf k}) = \frac{{\bf k}\times {\bf i_{v,b,cr}}}{|{\bf k}\times {\bf i_{v,b,cr}}|},   \eqno{(16)}
$$
${\bf i_{v,b,cr}}$ is random unit vector (unique for each subscript ${\bf v}, {\bf b}, {\bf cr}$) and updated whenever the forcing is initialized, $\varepsilon_{s} $ is total power of energy sources, $C=\varepsilon_{cr}/\varepsilon_{s}$ is level of injection of relative cross-helicity ($\varepsilon_{cr}$ is power of the sources of the cross-helicity). The large-scale forcing was applied in the wavenumber range $1 \leq k \leq 3$.

   Figures 2 and 3 show spectra of magnetic energy observed in the DNS reported in the Ref. \cite{tit} for $C=0.3$ and $C=0.6$ respectively (the spectral data were taken from the Fig. 2b of the Ref. \cite{tit}). The Reynolds and magnetic Reynolds numbers $Re =Re_m = 2094$. The dashed curves indicate correspondence to the stretched exponential spectrum Eq. (9).\\

   It is also interesting to consider a free (without forcing) decaying MHD turbulence with initially strong cross-correlation. Results of a DNS of such kind were reported in the Ref. \cite{baer}, where the so-called 3D Orszag-Tang flow \cite{spb} (with equal initial kinetic and magnetic energy and with the initial global cross-helicity normalized by total energy equal to 0.405) was taken as an initial condition. 
   
     Figures 4 and 5 show spectra of magnetic energy observed in this DNS for the time of decay $t=1$ and $t=10$ (in the DNS terms) respectively. The spectral data were taken from Fig. 12 of the Ref. \cite{spb}. The dashed curves indicate correspondence to the stretched exponential spectrum Eq. (9).
\begin{figure} \vspace{-1.5cm}\centering
\epsfig{width=.45\textwidth,file=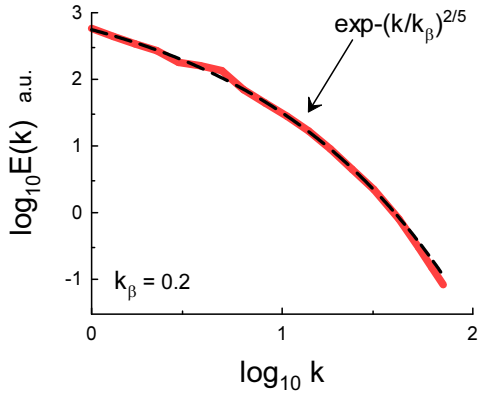} \vspace{-4.3cm}
\caption{Magnetic energy spectrum for $t=1$.} 
\end{figure}

\begin{figure} \vspace{-1.9cm}\centering
\epsfig{width=.45\textwidth,file=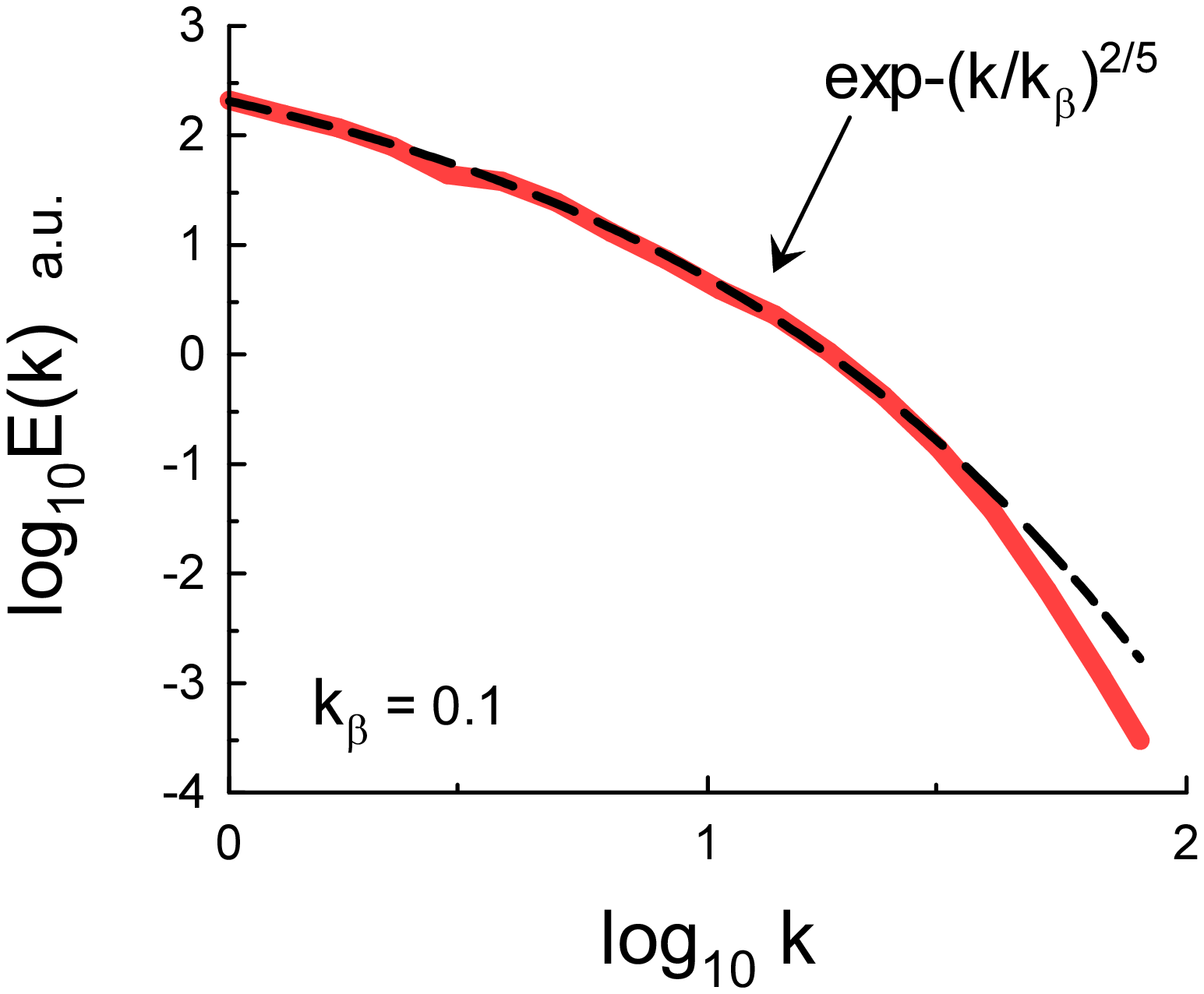} \vspace{-3.9cm}
\caption{Magnetic energy spectrum for $t=10$.} 
\end{figure}
\begin{figure} \vspace{-0.4cm}\centering
\epsfig{width=.45\textwidth,file=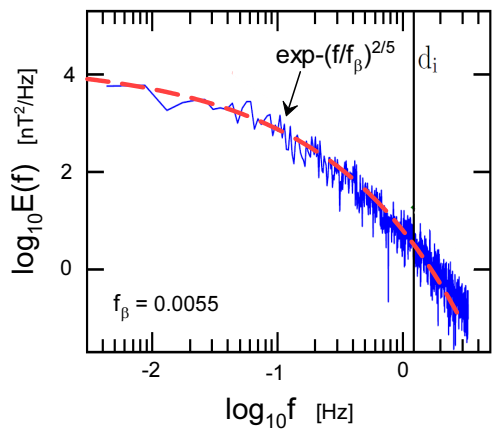} \vspace{-3.9cm}
\caption{Magnetic energy spectrum in the Earth's magnetosheath.} 
\end{figure}

\section{IV. Cross-helicity in the Earth's Magnetosheath}   

  The magnetohydrodynamic description is considered as an adequate one for the large-scale processes in the solar wind and in the Earth's magnetospheric region located downstream of the bow shock (the so-called Earth's magnetosheath). In the recent Ref. \cite{ban} the magnetic energy spectrum was computed using data obtained by the Magnetospheric Multiscale  (MMS) Mission spacecraft operated in the Earth's magnetosheath \cite{bur}. The low Mach number and density fluctuations (Table 1 of the Ref. \cite{ban}) justify applicability of the incompressible magnetohydrodynamics for the large scales. The estimated by the authors normalized cross-helicity $\sigma_c \simeq 0.24$ (Table 2 of the Ref. \cite{ban}) was smaller than those which can be observed in the solar wind but was certainly not negligible (let us recall that maximal value of $\sigma_c =1$).
  
    Figure 6 shows the magnetic energy spectrum against frequency (the spectral data were taken from the Fig. 2 of the Ref. \cite{ban}). Since in this case the wind's mean velocity $|\langle {\bf V} \rangle|$ in the spacecraft frame is considerably larger than the representative velocity fluctuations the Taylor "frozen" hypothesis $k \simeq 2\pi f/ |\langle {\bf V}| \rangle$ can be applied (see, for instance, Refs. \cite{hb},\cite{tbn} and references therein). Following to this hypothesis the temporal dynamics measured by the probe merely reflects the spatial one convected past the probe by the mean (nearly constant) velocity. Therefore, it is not a true frequency spectrum but actually a wavenumber spectrum (about intrinsic time dependence of the magnetic field see the Section VIII). Hence, the dashed curve in the Fig. 6 indicates correspondence to the stretched exponential {\it wavenumber} spectrum Eq. (9) and the scale $k_{\beta}\simeq 2\pi f_{\beta}/ |\langle {\bf V} \rangle|$ (it is clear that the $k_{\beta}$ corresponds to large-scale spatial structures in this case). The vertical black bar indicates the ion inertial length $d_i$ ($kd_i=1$).

\section{V. Cross-helicity vs. total energy}

   It was already mentioned in the Introduction that since the non-zero cross-helicity is related to lack of the reflection symmetry (unlike the velocity field ${\bf v}$, which is a polar vector, the magnetic field ${\bf B}$ is an axial vector) the $H_{cr}$ and $h_{cr}$ Eqs. (1-2) can be non-zero only when the global reflection symmetry is broken. Their role in MHD is similar to the role of the ideal invariants based on the hydrodynamic helicity $({\bf v} \cdot  {\boldsymbol \omega})$ (where ${\boldsymbol \omega} = \nabla \times {\bf v}$ is vorticity) in ordinary hydrodynamics (see the Refs. \cite{lt},\cite{b1} and for reviews the Refs. \cite{l},\cite{mt} and references therein). In particular, in the case of global reflection symmetry $H_{cr}=h_{cr}=0$. However, the higher (even) moments of the cross-helicity fluctuations can nevertheless be finite and constant, due to spatially localized lack of the reflection symmetry (with mutual compensation of contribution of the spatial areas with different sign of the cross-helicity distribution into the global cross-helicty). Therefore, the higher moments are of a special interest. Moreover, even in 
the cases of lack of the global reflection symmetry (when $H_{cr}$ and $h_{cr}$ are non-zero) the second moment can dominate extended inertial range of the total energy spectrum (as it will be shown below, cf. also the Ref. \cite{b1}).  \\ 
\begin{figure} \vspace{-1.7cm}\centering
\epsfig{width=.45\textwidth,file=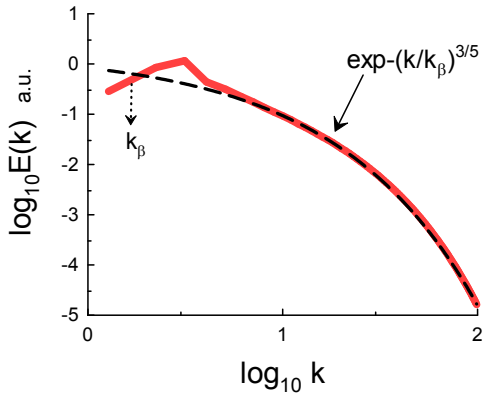} \vspace{-4.1cm}
\caption{Total energy spectrum for $C=0.3$.}
\end{figure}
\begin{figure} \vspace{-1.85cm}\centering
\epsfig{width=.44\textwidth,file=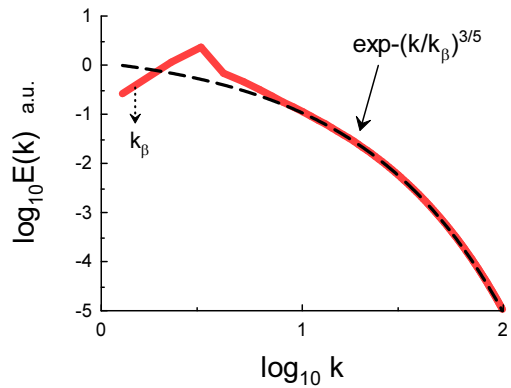} \vspace{-3.7cm}
\caption{Total energy spectrum for $C=0.6$.} 
\end{figure}

  Since dynamics of the magnetic field ${\bf b}$ in the non-dissipative case 
$$
\frac{\partial {\bf b}}{\partial t} = \nabla \times ( {\bf v} \times
    {\bf b})   \eqno{(17)}
$$    
has the same form as for vorticity ${\boldsymbol \omega}$ (the initial conditions for this equation, of course, belong to a much wider class than those allowed for the vorticity's dynamical equation \cite{mt}) the same consideration that was applied to the helicity moments \cite{lt},\cite{mt} can be applied to the cross-helicity moments as well. Namely, let us divide the entire volume of motion into cells $V_j$ with boundary conditions ${\bf b} \cdot {\bf n}=0$ on the bounding surfaces $S_j$ moving with the fluid. Then the 'localized' in the volume $V_j$ cross-helicity 
$$
H_{cr,j} = \int_{V_j} {\bf v}({\bf x},t)\cdot {\bf b}({\bf x},t) d{\bf r}  \eqno{(18)}
$$
is a non-dissipative (ideal) invariant of the motion. The second order moment of the cross-helicity distribution in this case can be defined as 
$$
I_{cr} = \lim_{V \rightarrow  \infty} \frac{1}{V} \sum_j H_{cr,j}^2  \eqno{(19)}
$$
and it is, due to its construction, a non-dissipative (ideal) invariant of the motion (it is interesting to compare this approach to the multifractal one, see for instance Ref. \cite{bt} and references therein). \\

  Let us consider a characteristic 'velocity' - $\tilde{v}_c$  defined for the total energy:
$$
{\mathcal E} = \frac{1}{2}({\bf v}^2 + {\bf b}^2)     \eqno{(20})
$$
(in the Alfv\'enic units the normalized magnetic field ${\bf b} = {\bf B}/\sqrt{\mu_0\rho}$ have the same dimension as velocity). From the dimensional considerations we obtain
$$
\tilde{v}_c \propto I_{cr}^{1/4} k_c^{3/4},  \eqno{(21)}
$$
i.e. $\alpha =3/4$ in this case (cf. Eq. (6)). Then from the Eq. (7) we obtain $\beta =3/5$, i.e. the total energy spectrum 
$$
E(k) \propto \exp-(k/k_{\beta})^{3/5}.   \eqno{(22)}
$$

\section{VI. Direct numerical simulations - II}

   Let us again start from the direct numerical simulation described in the Ref. \cite{tit} (see the Section: "Direct numerical simulations - I" above, Figs. (2-3)), but now we will consider results obtained for the total energy spectrum and reported in recent Ref. \cite{mag}. The {\it random} large-scale forcing terms Eqs. (14-15) were applied in the wavenumber range $1 \leq k \leq 3$ and the Kolmogorov (dissipation) wavenumber $k_K = k_m \simeq 100$ ($Re = Re_m \simeq 2094$). Figures 7 and 8 show spectra of total energy for $C=0.3$ and $C=0.6$ respectively (the spectra correspond to the Fig. 1a of the Ref. \cite{mag}). The dashed curves indicate correspondence to the stretched exponential spectrum Eq. (22) and the dotted arrows indicate position of the scale $k_{\beta}$. 

   One can see that the extended inertial range penetrates rather deep into the near dissipation range. This phenomenon can be related to the coherency with the large scales introduced by domination of the distributed chaos (cf. position of the scale $k_{\beta}$ in the Figs. (7-8)). It should be noted in this respect that the cross helicity is known as a factor that introduce a nonlocal interaction between the small- and large-scale fluctuations. The coherency results in the adiabatic invariance of the $I_{cr}$ up to the the dissipation scale (cf. also Refs. \cite{lst},\cite{mof}).  \\

   Let us also consider dynamics of the total energy in the freely decaying MHD turbulence with initially strong cross-correlation generated by the initial conditions taken in the form of the 3D Orszag-Tang flow (cf. the Section: "Direct numerical simulations - I" above, Figs. (4-5)). Figures 9 and 10 show spectra of total energy for the Taylor-Reynolds numbers $Re_{\lambda} = 45$ and  $Re_{\lambda} = 30$ respectively (the spectral data were taken from the Fig. 2a of the Ref. \cite{gibbon} and correspond to the early times of the decay, the $Re_{\lambda}$ is decreasing with time of decay). The magnetic Prandtl number $P_m$ was again taken equal to 1, i.e $\nu=\eta$. The dashed curves indicate correspondence to the stretched exponential spectrum Eq. (22) and the dotted arrows indicate position of the scale $k_{\beta}$. \\

   Finally, let us consider results of DNS with external/mean magnetic field reported in Ref. \cite{cho} (this case is  important for astrophysics). The imposed magnetic field was a moderate one $b_0 =1$ (in the terms of the DNS \cite{cho}). However, the statistically stationary MHD turbulence is anisotropic. The MHD turbulence was incompressible and was described by the Eq. (10-12) with $\nu=\eta$ and random isotropic forcing ${\bf f_v}$ (whereas ${\bf f_b} =0$). The large-scale random energy injection has its peak at $k \simeq 2.5$.
\begin{figure} \vspace{-1.5cm}\centering
\epsfig{width=.42\textwidth,file=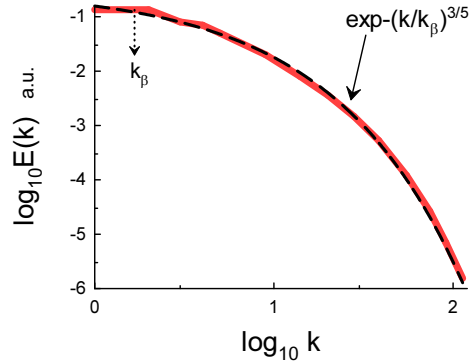} \vspace{-4cm}
\caption{Total energy spectrum for $Re_{\lambda} = 45$.} 
\end{figure}
\begin{figure} \vspace{-0.5cm}\centering
\epsfig{width=.45\textwidth,file=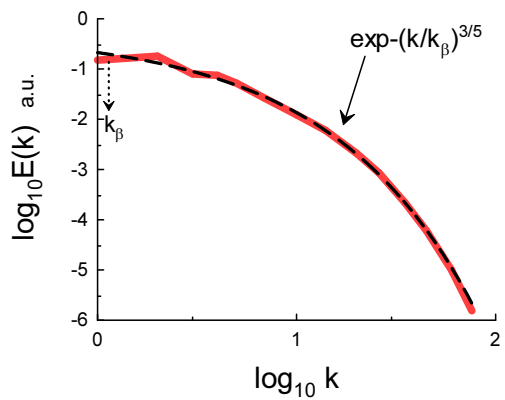} \vspace{-4.2cm}
\caption{Total energy spectrum for $Re_{\lambda} = 30$.} 
\end{figure}
\begin{figure} \vspace{-1.5cm}\centering
\epsfig{width=.45\textwidth,file=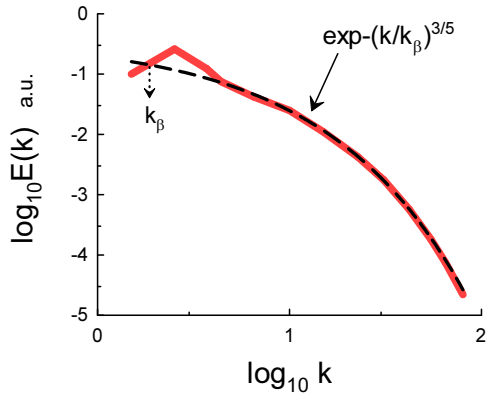} \vspace{-4.2cm}
\caption{Total energy spectrum for $b_0 =1$.} 
\end{figure}
\begin{figure} \vspace{-0.5cm}\centering
\epsfig{width=.445\textwidth,file=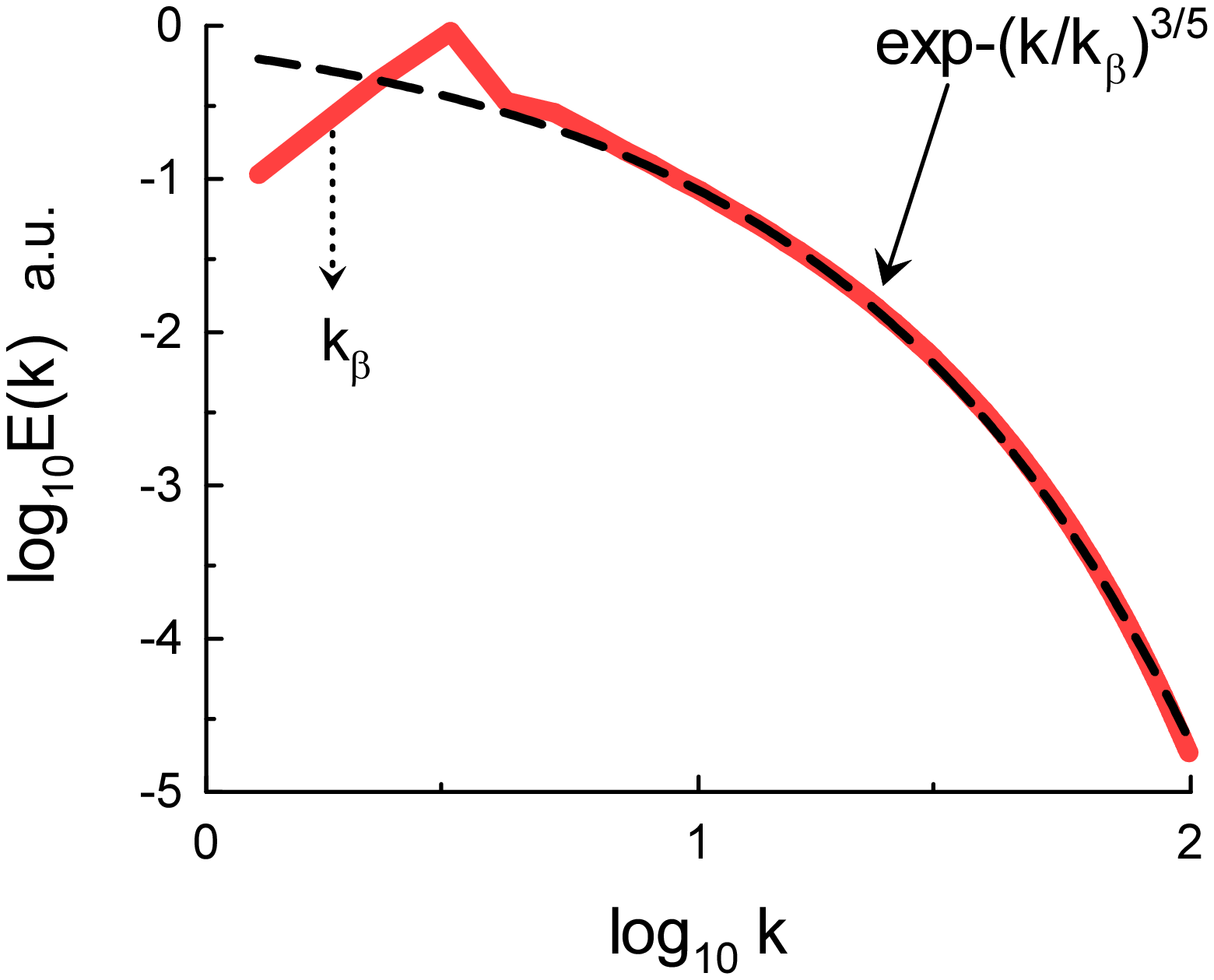} \vspace{-3.85cm}
\caption{Total energy spectrum for $C = 0$.} 
\end{figure}
\begin{figure} \vspace{-0.5cm}\centering
\epsfig{width=.445\textwidth,file=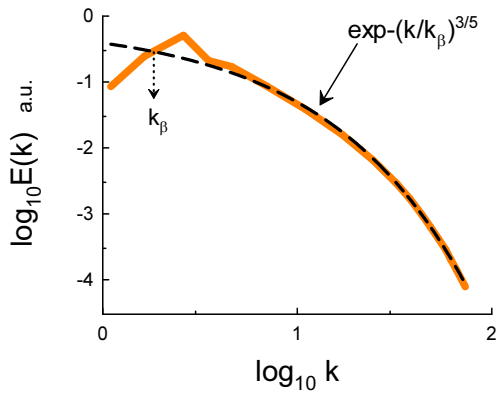} \vspace{-4.5cm}
\caption{Magnetic energy spectrum for $C = 0$.} 
\end{figure}
   Figure 11 shows spectrum of total energy observed in this DNS for $b_0 =1$ (the spectral data were taken from the Fig. 16 of the Ref. \cite{cho}). The dashed curve indicates correspondence to the stretched exponential spectrum Eq. (22) and the dotted arrows indicate position of the scale $k_{\beta}$.

\section{VII. Spontaneous breaking of local reflection symmetry}

The cross-helicity density changes its sign at mirror reflection of the coordinate system (a pseudoscalar). Therefore, the global quantities  Eqs. (1-2) are non-zero only if there is a global lack of reflection (mirror) symmetry, and the non-zero average cross-helicity indicates the breakage of global reflection (mirror) symmetry.  However, even when the average cross-helicity is negligible the magnitude of its density can be locally large (in the vicinity of current and vorticity sheets, for instance, see the Rerfs. \cite{mene},\cite{matt2} and references therein). After the global averaging over the localized patches with negative and positive cross-helicity one has the average cross-helicity close to zero (there exits a view that the MHD turbulence with close to zero average cross-helicity can be generally considered as a superposition of these patches, see for instance Ref. \cite{pb} and references therein). Moreover, the patches can possess a kind of hierarchical structure. Namely, inside the patches there can exist smaller ones with more strong chirality (of different signs) and so on \cite{pb}. The above considered (adiabatic) invariant $I_{cr}$ Eq. (19) is rather adequate tool for description of this situation. In particular, as it was already mentioned, this (adiabatic) invariant can be finite even in the cases when the average cross-helicty is close to zero. \\

  To check applicability of this consideration let us return to the results obtained in the direct numerical simulation reported in the Ref. \cite{mag} (see previous Section) but now for $C = 0$, i.e. for the case with negligible average cross-helicity. Figure 12 shows spectrum of total energy obtained for $C=0$ (the spectrum corresponds to the Fig. 1a of the Ref. \cite{mag}). The dashed curve indicates correspondence to the stretched exponential spectrum Eq. (22) and the dotted arrow indicates position of the scale $k_{\beta}$. \\

  It is instructive to compare Fig. 12 with the Figs. (7-8) showing the total energy spectra for $C=0.3$ and $C=0.6$ (corresponding to considerable average cross-helicity).  This comparison allows us to conclude that the effect of the spontaneous breaking of the local reflection (mirror) symmetry indeed takes place in this case and can be described in the terms of the (adiabatic) invariant $I_{cr}$ (see also below).
  
    It should be noted that in the Alfv\'enic units the normalized magnetic field ${\bf b} = {\bf B}/\sqrt{\mu_0\rho}$ has the same dimension as velocity and the Eq. (21) can be replaced by equation 
$$
b_c \propto I_{cr}^{1/4} k_c^{3/4}  \eqno{(23)}
$$   
Therefore, unlike the cases with considerable average cross-helicty (see Figs. 2 and 3), for $C=0$ the {\it magnetic} energy spectrum has behaviour similar to that shown in the Fig. 12. Figure 13 shows the magnetic energy spectrum obtained for $C=0$ (the spectral data for the Fig. 13 were taken from the Fig. 2b of the Ref. \cite{tit}, as for the Figs. 2 and 3). The dashed curve indicates correspondence to the stretched exponential spectrum Eq. (22), whereas for the Figs. 2 and 3 the magnetic energy spectrum corresponds to the Eq. (9). \\
  
    Generally the MHD turbulence exhibits a wide variety of energy spectra depending on initial-boundary conditions and type of forcing. At present time the necessary and sufficient conditions for the spontaneous breaking of local reflection symmetry in MHD turbulence are not known.

\section{VIII. Spatio-temporal distributed chaos}
 
  Is the above considered distributed MHD chaos exclusively spatial or is it a spatio-temporal one? To answer this question let us replace the spatial (wavenumber)  Eq. (21) by its temporal (frequency) analogue using the dimensional considerations:
$$
\tilde{v}_c \propto I_{cr}^{1/7} f_c^{3/7}  \eqno{(24)}
$$
where $f_c$ is a characteristic frequency. It follows from the Eq. (24) that $\alpha =3/7$. Then from the Eq. (7) we obtain $\beta =6/13$, i.e. the frequency total energy spectrum is 
$$
E(f) \propto \exp-(f/f_{\beta})^{6/13}.   \eqno{(25)}
$$   
in this case.
  
   In the Alfv\'enic units the normalized magnetic field ${\bf b} = {\bf B}/\sqrt{\mu_0\rho}$ has the same dimension as velocity and the Eq. (24) can be replaced by equation 
$$
b_c \propto I_{cr}^{1/7} f_c^{3/7},  \eqno{(26)}
$$    
i.e. in this case the magnetic energy spectrum has the same form as the total energy spectrum Eq. (25). \\

  In Ref. \cite{dm} results of a direct numerical simulation of the isotropic homogeneous MHD turbulence of incompressible conducting fluid were reported and an Eulerian {\it frequency} spectrum of magnetic energy was constructed using 64 point-like probes set in a middle plane of 3D spatial box in a regular  $8\times8$ array (the Reynolds and magnetic Reynolds numbers were $Re=Re_m=400$). The long time series obtained with these probes were used to compute the average (over all probes) frequency spectrum of magnetic energy fluctuations. The initial state for this simulation was a random phased set of velocity and magnetic filed fluctuations (in equipartition) in the shell $1 \leq |{\bf k}| \leq 4$. The initial average cross-helicity was negligible in comparison with the energy. Due to the {\it uncorrelated} character of the random forcing ${\bf f_v}$ and ${\bf f_b}$ (in the range $1 \leq  k \leq 2$) there was no statistical injection of cross-helicity and the average cross-helicity was negligible also in the steady state of the MHD turbulence. Therefore, one can expect that the spontaneous breaking of local reflection symmetry can take place at this simulation. \\
 
\begin{figure} \vspace{-1.7cm}\centering
\epsfig{width=.445\textwidth,file=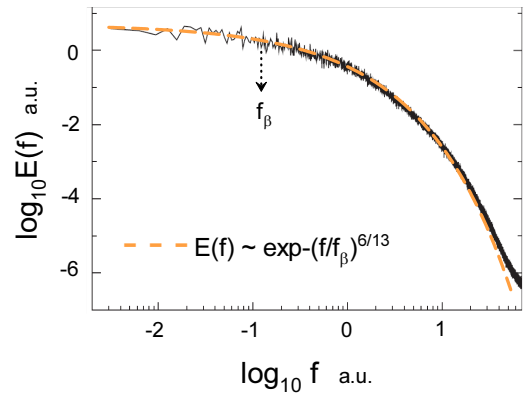} \vspace{-4.1cm}
\caption{Frequency spectrum of magnetic energy.} 
\end{figure}
   Figure 14 shows frequency spectrum of magnetic energy obtained in this DNS for the incompressible MHD turbulence without background magnetic field. The spectrum corresponds to the Fig. 2 (upper panel) of the Ref. \cite{dm}. The dashed curve indicates correspondence to the stretched exponential spectrum Eq. (25) and the dotted arrow indicates position of the scale $f_{\beta}$. One can see that the four decades range of the frequency spectrum supports the spatio-temporal character of the distributed MHD chaos and indicates the spontaneous breaking of the local reflection symmetry in this case (cf. Fig. 13 corresponding to the wavenumber spectrum).
 
\section{IX. Effects of mean magnetic filed} 

  We have already mentioned results of DNS with a moderate imposed magnetic field (see Fig. 11). Presence of a strong mean magnetic field can result in fundamental changes in the properties of MHD turbulence. Even when one ignore anisotropy introduced by the mean magnetic field (suggesting, for instance, the local isotropy), as it was made in the pioneering papers \cite{ir},\cite{kr}, the dimensional parameter governing the inertial range of scales in the Kolmogorov theory - $\varepsilon$, should be replaced by the parameter $\varepsilon b_0$ (where $\varepsilon$ is the energy dissipation rate and the normalised mean magnetic field $b_0 = B_0/\sqrt{\mu_0\rho}$ has the same dimension as velocity). A vigorous discussion about validity of this replacement in solar wind, for instance, still takes place in interpretation of scaling properties of the modern spacecraft data (spectra $E(k) \propto k^{-5/3}$ versus spectra $E(k) \propto k^{-3/2}$) and a few new theoretical approaches were suggested (let us mention the Refs. \cite{gs},\cite{bold}).\\
\begin{figure} \vspace{-1.7cm}\centering
\epsfig{width=.445\textwidth,file=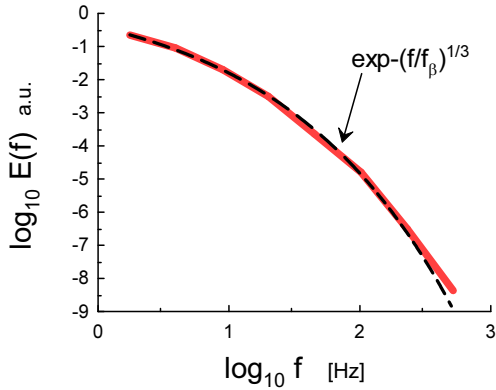} \vspace{-4.1cm}
\caption{Magnetic energy spectrum in the MHD wind (plasma) tunnel.} 
\end{figure}
  With the replacement $\varepsilon \rightarrow \varepsilon b_0$ the Eq. (8) should be replaced by the equation 
$$
B_c \propto    |\langle  {\bf v}\cdot {\bf B} \rangle|~(\varepsilon b_0)^{-1/4}~ k_c^{1/4} \eqno{(27)} 
$$
that gives $\alpha =1/4$ and, correspondingly, one obtains from the  Eq. (7) $\beta = 1/3$. Hence the magnetic energy spectrum is
$$
E(k) \propto \exp-(k/k_{\beta})^{1/3}   \eqno{(28)}
$$  
in this case. \\

   Recent paper Ref. \cite{sbl} reports results obtained in a laboratory experiment with magnetic turbulent plasma in a MHD wind tunnel. The experiment was especially designed in order to model solar wind magnetohydrodynamics. Figure 15 shows magnetic energy spectrum (ensemble averaged) measured in this experiment (the spectral data for this figure were taken from Fig. 15a of the Ref. \cite{sbl}). The Taylor hypothesis (see Section IV) relates the frequency spectrum to analogous spatial (wavenumber) spectrum $E(k)$ and the dashed curve indicates correspondence to the stretched exponential spectrum Eq. (28).\\
   
      As for the solar wind itself the spacecraft Ulysses and Helios-1 measurements, made at high and low heliolatitudes respectively, provide a general picture of its magnetohydrodynamics from the distances 4.5 AU (Ulysses) to 0.3 AU  (Helios-1) from the Sun. These measurements in the high-speed streams show a strong similarity for the Ulysses and Helios-1 data (see, for instance, Ref. \cite{hb} and references therein).\\

      Figure 16 shows an example of magnetic energy spectra obtained by magnetometers of the Helios-1. The spectral data for the Fig. 16 were taken from Fig. 3 of the Ref. \cite{tbn} (full speed mapping). The dashed curve indicates correspondence to the stretched exponential spectrum Eq. (28). The dotted arrow indicates position of the scale $k_{\beta}$. \\
  
  Figures 17 and 18 show magnetic energy spectrum corresponding to the Ulysses data for the period 1993-1996yy at high solar wind speed and at high heliolatitudes. The spectral data for these figures were taken from Fig. 3 of the Ref. \cite{bran} for $1.5 < R < 2.8$ AU and for $2.8 < R < 4.5$ AU, respectively ($R$ is distance between the spacecraft and the Sun). Before averaging over the data sets the spectra were rescaled by factor $4\pi R^2$. The dashed curves correspond to the Eq. (28).
\begin{figure} \vspace{-3.3cm}\centering
\epsfig{width=.445\textwidth,file=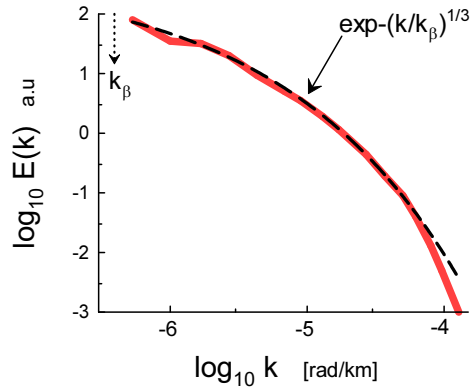} \vspace{-2.5cm}
\caption{Magnetic energy spectrum in the solar wind (the Helios-1 data).} 
\end{figure}

\begin{figure} \vspace{-0.3cm}\centering
\epsfig{width=.445\textwidth,file=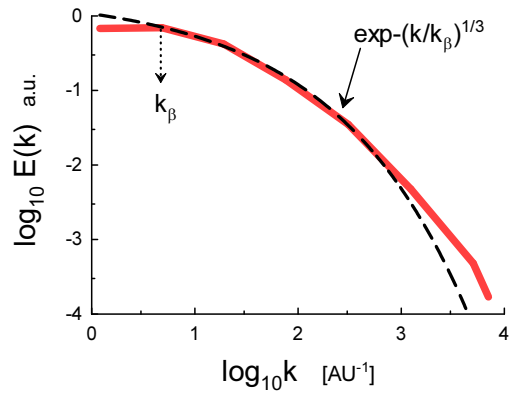} \vspace{-4.1cm}
\caption{Magnetic energy spectrum in the solar wind (the Ulysses data for $1.5 < R < 2.8$ AU).} 
\end{figure}
\begin{figure} \vspace{-1.5cm}\centering
\epsfig{width=.445\textwidth,file=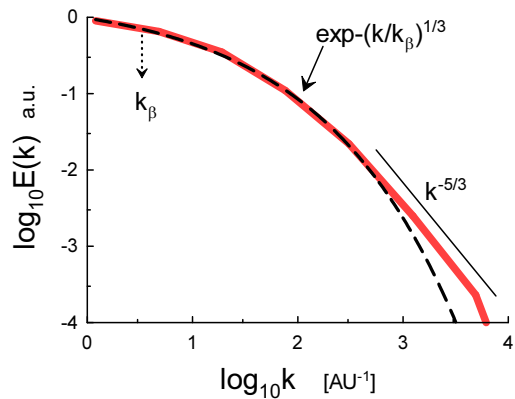} \vspace{-4.3cm}
\caption{Magnetic energy spectrum in the solar wind (the Ulysses data for $2.8 < R < 4.5$ AU).} 
\end{figure}

\section{Acknowledgement}

I thank  E. Levich for stimulating discussions, A. Beresnyak, V. Pipin, D.D. Sokoloff and J. Shebalin for comments,  
R. Grappin, R. Stepanov and V. Titov for sharing their data and additional information.

\end{document}